\documentclass[onecolumn,12pt,draftcls,peerreview]{IEEEtran}

\IEEEoverridecommandlockouts

\usepackage{amssymb, amsmath}
\usepackage{amsfonts}
\usepackage{bbm}
\usepackage{mathrsfs}
\usepackage{xspace}
\usepackage{bm}

\usepackage{cite}
\usepackage{epsfig}
\usepackage{cases}
\usepackage{psfrag}
\usepackage{enumerate}

\usepackage{makeidx}
\usepackage{supertabular}

\newtheorem{theorem}{Theorem}
\newtheorem{definition}{Definition}

\newtheorem{remark}{Remark}

\newenvironment{textbmatrix}{   \setlength{\arraycolsep}{2.5pt}%
                                                                \big[\begin{matrix}}{\end{matrix}\big]%
                                                                \raisebox{0.08ex}{\vphantom{M}}}

\def\be{\begin{equation}}
\def\ee{\end{equation}}
\def\een{\nonumber \end{equation}}
\def\mat{\begin{bmatrix}}
\def\emat{\end{bmatrix}}
\def\btm{\begin{textbmatrix}}
\def\etm{\end{textbmatrix}}

\def\ba#1\ea{\begin{align}#1\end{align}}
\def\bs#1\es{\begin{split}#1\end{split}} 
\def\bg#1\eg{\begin{gather}#1\end{gather}} 
\def\bi#1\ei{\begin{itemize}#1\end{itemize}}

\newcommand{\safemath}[2]{\newcommand{#1}{\ensuremath{#2}\xspace}}

\DeclareMathOperator{\diag}{diag}                       
\DeclareMathOperator*{\argmax}{arg\;max}                
\DeclareMathOperator{\Exop}{\mathbb{E}}         

\safemath{\interior}{\mathrm{Int}}                       

\safemath{\dfn}{:=}                                                     
\safemath{\dirac}{\delta}                                       

\safemath{\SNR}{\text{\sc snr}}                                 
\safemath{\No}{N_0}                                                     
\safemath{\Es}{E_s}                                                     
\safemath{\Eb}{E_b}                                                     
\safemath{\EbNo}{\frac{\Eb}{\No}}
\safemath{\EsNo}{\frac{\Es}{\No}}

\DeclareMathOperator{\CHop}{\ensuremath{\mathbb{H}}} 
\safemath{\tvir}{h_{\CHop}}                                     
\safemath{\tvtf}{L_{\CHop}}                                     
\safemath{\spf}{S_{\CHop}}                                              
\safemath{\bff}{H_{\CHop}}                                      

\safemath{\ircf}{R_{h}}                                         
\safemath{\scf}{R_{S}}                                          
\safemath{\tfcf}{R_{L}}                                         
\safemath{\bfcf}{R_{H}}                                         

\safemath{\mi}{I}                                                       
\safemath{\capacity}{C}                                         

\newcommand{\pdf[2]}{f_{{#1}}\left({#2}\right)}                    

\safemath{\uniform}{\mathcal{U}}                        
\safemath{\normal}{\mathcal{N}}                         
\safemath{\circnorm}{\mathcal{CN}}                      
\safemath{\mchain}{\leftrightarrow}                     

\safemath{\dB}{\,\mathrm{dB}}
\safemath{\dBm}{\,\mathrm{dBm}}
\safemath{\Hz}{\,\mathrm{Hz}}
\safemath{\kHz}{\,\mathrm{kHz}}
\safemath{\MHz}{\,\mathrm{MHz}}
\safemath{\GHz}{\,\mathrm{GHz}}
\safemath{\s}{\,\mathrm{s}}
\safemath{\ms}{\,\mathrm{ms}}
\safemath{\mus}{\,\mathrm{\mu s}}
\safemath{\ns}{\,\mathrm{ns}}
\safemath{\meter}{\,\mathrm{m}}
\safemath{\km}{\,\mathrm{km}}
\safemath{\mm}{\,\mathrm{mm}}
\safemath{\cm}{\,\mathrm{cm}}
\safemath{\m}{\,\mathrm{m}}
\safemath{\W}{\,\mathrm{W}}
\safemath{\J}{\,\mathrm{J}}
\safemath{\K}{\,\mathrm{K}}
\safemath{\bit}{\,\mathrm{bit}}
\safemath{\nW}{\,\mathrm{nW}}
\safemath{\muW}{\,\mathrm{$\mu$W}}
\safemath{\Watt}{\,\mathrm{W}}

\safemath{\define}{\triangleq}                  

\safemath{\equivalent}{\sim}
\safemath{\distas}{\sim}                                        

\safemath{\reals}{\mathbb{R}}
\safemath{\positivereals}{\mathbb{R}^{+}}
\safemath{\integers}{\mathbb{Z}}
\safemath{\posint}{\mathbb{Z}_{+}}
\safemath{\naturals}{\mathbb{N}}
\safemath{\complexset}{\mathbb{C}}

\safemath{\setA}{\mathcal{A}}
\safemath{\setB}{\mathcal{B}}
\safemath{\setC}{\mathcal{C}}
\safemath{\setD}{\mathcal{D}}
\safemath{\setE}{\mathcal{E}}
\safemath{\setF}{\mathcal{F}}
\safemath{\setG}{\mathcal{G}}
\safemath{\setH}{\mathcal{H}}
\safemath{\setI}{\mathcal{I}}
\safemath{\setJ}{\mathcal{J}}
\safemath{\setK}{\mathcal{K}}
\safemath{\setL}{\mathcal{L}}
\safemath{\setM}{\mathcal{M}}
\safemath{\setN}{\mathcal{N}}
\safemath{\setO}{\mathcal{O}}
\safemath{\setP}{\mathcal{P}}
\safemath{\setQ}{\mathcal{Q}}
\safemath{\setR}{\mathcal{R}}
\safemath{\setS}{\mathcal{S}}
\safemath{\setT}{\mathcal{T}}
\safemath{\setU}{\mathcal{U}}
\safemath{\setV}{\mathcal{V}}
\safemath{\setW}{\mathcal{W}}
\safemath{\setX}{\mathcal{X}}
\safemath{\setY}{\mathcal{Y}}
\safemath{\setZ}{\mathcal{Z}}
\safemath{\emptySet}{\varnothing}

\safemath{\bma}{\mathbf{a}}
\safemath{\bmb}{\mathbf{b}}
\safemath{\bmc}{\mathbf{c}}
\safemath{\bmd}{\mathbf{d}}
\safemath{\bme}{\mathbf{e}}
\safemath{\bmf}{\mathbf{f}}
\safemath{\bmg}{\mathbf{g}}
\safemath{\bmh}{\mathbf{h}}
\safemath{\bmi}{\mathbf{i}}
\safemath{\bmj}{\mathbf{j}}
\safemath{\bmk}{\mathbf{k}}
\safemath{\bml}{\mathbf{l}}
\safemath{\bmm}{\mathbf{m}}
\safemath{\bmn}{\mathbf{n}}
\safemath{\bmo}{\mathbf{o}}
\safemath{\bmp}{\mathbf{p}}
\safemath{\bmq}{\mathbf{q}}
\safemath{\bmr}{\mathbf{r}}
\safemath{\bms}{\mathbf{s}}
\safemath{\bmt}{\mathbf{t}}
\safemath{\bmu}{\mathbf{u}}
\safemath{\bmv}{\mathbf{v}}
\safemath{\bmw}{\mathbf{w}}
\safemath{\bmx}{\mathbf{x}}
\safemath{\bmy}{\mathbf{y}}
\safemath{\bmz}{\mathbf{z}}

\bmdefine{\biad}{a}
\bmdefine{\bibd}{b}
\bmdefine{\bicd}{c}
\bmdefine{\bidd}{d}
\bmdefine{\bied}{e}
\bmdefine{\bifd}{f}
\bmdefine{\bigd}{g}
\bmdefine{\bihd}{h}
\bmdefine{\biid}{i}
\bmdefine{\bijd}{j}
\bmdefine{\bikd}{k}
\bmdefine{\bild}{l}
\bmdefine{\bimd}{m}
\bmdefine{\bind}{n}
\bmdefine{\biod}{o}
\bmdefine{\bipd}{p}
\bmdefine{\biqd}{q}
\bmdefine{\bird}{r}
\bmdefine{\bisd}{s}
\bmdefine{\bitd}{t}
\bmdefine{\biud}{u}
\bmdefine{\bivd}{v}
\bmdefine{\biwd}{w}
\bmdefine{\bixd}{x}
\bmdefine{\biyd}{y}
\bmdefine{\bizd}{z}

\bmdefine{\bixid}{\xi}
\bmdefine{\bilambdad}{\lambda}
\bmdefine{\bimud}{\mu}
\bmdefine{\bithetad}{\theta}
\bmdefine{\biphid}{\phi}

\safemath{\bmia}{\biad}
\safemath{\bmib}{\bibd}
\safemath{\bmic}{\bicd}
\safemath{\bmid}{\bidd}
\safemath{\bmie}{\bied}
\safemath{\bmif}{\bifd}
\safemath{\bmig}{\bigd}
\safemath{\bmih}{\bihd}
\safemath{\bmii}{\biid}
\safemath{\bmij}{\bijd}
\safemath{\bmik}{\bikd}
\safemath{\bmil}{\bild}
\safemath{\bmim}{\bimd}
\safemath{\bmin}{\bind}
\safemath{\bmio}{\biod}
\safemath{\bmip}{\bipd}
\safemath{\bmiq}{\biqd}
\safemath{\bmir}{\bird}
\safemath{\bmis}{\bisd}
\safemath{\bmit}{\bitd}
\safemath{\bmiu}{\biud}
\safemath{\bmiv}{\bivd}
\safemath{\bmiw}{\biwd}
\safemath{\bmix}{\bixd}
\safemath{\bmiy}{\biyd}
\safemath{\bmiz}{\bizd}

\safemath{\bmxi}{\bixid}
\safemath{\bmlambda}{\bilambdad}
\safemath{\bmmu}{\bimud}
\safemath{\bmtheta}{\bithetad}
\safemath{\bmphi}{\biphid}

\safemath{\bA}{\mathbf{A}}
\safemath{\bB}{\mathbf{B}}
\safemath{\bC}{\mathbf{C}}
\safemath{\bD}{\mathbf{D}}
\safemath{\bE}{\mathbf{E}}
\safemath{\bF}{\mathbf{F}}
\safemath{\bG}{\mathbf{G}}
\safemath{\bH}{\mathbf{H}}
\safemath{\bI}{\mathbf{I}}
\safemath{\bJ}{\mathbf{J}}
\safemath{\bK}{\mathbf{K}}
\safemath{\bL}{\mathbf{L}}
\safemath{\bM}{\mathbf{M}}
\safemath{\bN}{\mathbf{N}}
\safemath{\bO}{\mathbf{O}}
\safemath{\bP}{\mathbf{P}}
\safemath{\bQ}{\mathbf{Q}}
\safemath{\bR}{\mathbf{R}}
\safemath{\bS}{\mathbf{S}}
\safemath{\bT}{\mathbf{T}}
\safemath{\bU}{\mathbf{U}}
\safemath{\bV}{\mathbf{V}}
\safemath{\bW}{\mathbf{W}}
\safemath{\bX}{\mathbf{X}}
\safemath{\bY}{\mathbf{Y}}
\safemath{\bZ}{\mathbf{Z}}

\safemath{\bZero}{\mathbf{0}}

\bmdefine{\biAd}{A}
\bmdefine{\biBd}{B}
\bmdefine{\biCd}{C}
\bmdefine{\biDd}{D}
\bmdefine{\biEd}{E}
\bmdefine{\biFd}{F}
\bmdefine{\biGd}{G}
\bmdefine{\biHd}{H}
\bmdefine{\biId}{I}
\bmdefine{\biJd}{J}
\bmdefine{\biKd}{K}
\bmdefine{\biLd}{L}
\bmdefine{\biMd}{M}
\bmdefine{\biOd}{N}
\bmdefine{\biPd}{O}
\bmdefine{\biQd}{P}
\bmdefine{\biRd}{R}
\bmdefine{\biSd}{S}
\bmdefine{\biTd}{T}
\bmdefine{\biUd}{U}
\bmdefine{\biVd}{V}
\bmdefine{\biWd}{W}
\bmdefine{\biXd}{X}
\bmdefine{\biYd}{Y}
\bmdefine{\biZd}{Z}

\bmdefine{\biDelta}{\Delta}
\bmdefine{\biLambda}{\Lambda}
\bmdefine{\biPhi}{\Phi}
\bmdefine{\biSigma}{\Sigma}
\bmdefine{\biOmega}{\Omega}
\bmdefine{\biTheta}{\Theta}

\safemath{\bimA}{\biAd}
\safemath{\bimB}{\biBd}
\safemath{\bimC}{\biCd}
\safemath{\bimD}{\biDd}
\safemath{\bimE}{\biEd}
\safemath{\bimF}{\biFd}
\safemath{\bimG}{\biGd}
\safemath{\bimH}{\biHd}
\safemath{\bimI}{\biId}
\safemath{\bimJ}{\biJd}
\safemath{\bimK}{\biKd}
\safemath{\bimL}{\biLd}
\safemath{\bimM}{\biMd}
\safemath{\bimN}{\biNd}
\safemath{\bimO}{\biOd}
\safemath{\bimP}{\biPd}
\safemath{\bimQ}{\biQd}
\safemath{\bimR}{\biRd}
\safemath{\bimS}{\biSd}
\safemath{\bimT}{\biTd}
\safemath{\bimU}{\biUd}
\safemath{\bimV}{\biVd}
\safemath{\bimW}{\biWd}
\safemath{\bimX}{\biXd}
\safemath{\bimY}{\biYd}
\safemath{\bimZ}{\biZd}

\safemath{\bDelta}{\bielta}
\safemath{\bLambda}{\biLambda}
\safemath{\bPhi}{\biPhi}
\safemath{\bSigma}{\biSigma}
\safemath{\bOmega}{\biOmega}
\safemath{\bTheta}{\biTheta}

\safemath{\veca}{\bma}
\safemath{\vecb}{\bmb}
\safemath{\vecc}{\bmc}
\safemath{\vecd}{\bmd}
\safemath{\vece}{\bme}
\safemath{\vecf}{\bmf}
\safemath{\vecg}{\bmg}
\safemath{\vech}{\bmh}
\safemath{\veci}{\bmi}
\safemath{\vecj}{\bmj}
\safemath{\veck}{\bmk}
\safemath{\vecl}{\bml}
\safemath{\vecm}{\bmm}
\safemath{\vecn}{\bmn}
\safemath{\veco}{\bmo}
\safemath{\vecp}{\bmp}
\safemath{\vecq}{\bmq}
\safemath{\vecr}{\bmr}
\safemath{\vecs}{\bms}
\safemath{\vect}{\bmt}
\safemath{\vecu}{\bmu}
\safemath{\vecv}{\bmv}
\safemath{\vecw}{\bmw}
\safemath{\vecx}{\bmx}
\safemath{\vecy}{\bmy}
\safemath{\vecz}{\bmz}
\safemath{\vecZero}{\bZero}

\safemath{\vecxi}{\bmxi}
\safemath{\veclambda}{\bmlambda}
\safemath{\vecmu}{\bmmu}
\safemath{\vectheta}{\bmtheta}
\safemath{\vecphi}{\bmphi}

\safemath{\matA}{\bA}
\safemath{\matB}{\bB}
\safemath{\matC}{\bC}
\safemath{\matD}{\bD}
\safemath{\matE}{\bE}
\safemath{\matF}{\bF}
\safemath{\matG}{\bG}
\safemath{\matH}{\bH}
\safemath{\matI}{\bI}
\safemath{\matJ}{\bJ}
\safemath{\matK}{\bK}
\safemath{\matL}{\bL}
\safemath{\matM}{\bM}
\safemath{\matN}{\bN}
\safemath{\matO}{\bO}
\safemath{\matP}{\bP}
\safemath{\matQ}{\bQ}
\safemath{\matR}{\bR}
\safemath{\matS}{\bS}
\safemath{\matT}{\bT}
\safemath{\matU}{\bU}
\safemath{\matV}{\bV}
\safemath{\matW}{\bW}
\safemath{\matX}{\bX}
\safemath{\matY}{\bY}
\safemath{\matZ}{\bZ}
\safemath{\matZero}{\bZero}

\safemath{\matDelta}{\bDelta}
\safemath{\matLambda}{\bLambda}
\safemath{\matPhi}{\bPhi}
\safemath{\matSigma}{\bSigma}
\safemath{\matOmega}{\bOmega}
\safemath{\matTheta}{\bTheta}

\safemath{\matIdentity}{\matI}

\usepackage[usenames,dvipsnames]{color}

\newcommand{\sectionname}{Section}

\renewcommand{\figurename}{Fig.}

\newcommand{\theoremname}{Theorem}

\newcommand{\appendicesname}{Appendices}

\usepackage{algorithm,algorithmic}

\safemath{\crb}{\mathsf{CRB}}
\safemath{\mse}{\mathsf{MSE}}
\safemath{\var}{\mathsf{var}}
\safemath{\nvar}{\mathsf{nvar}}

\newcommand{\db[1]}{{#1}|_{\textrm{dB}}}

\begin{document}

\title{{Energy-Efficient Power Control for Contention- Based Synchronization in OFDMA Systems with Discrete Powers and Limited Feedback}}
\author{Giacomo~Bacci, \emph{Member, IEEE},
\thanks{The research leading to these results has received funding from the People Programme (Marie Curie Actions) of the European Union's Seventh Framework Programme (FP7/2007-2013) under REA grant agreement n.~PIOF-GA-2011-302520 GRAND-CRU ``Game-theoretic Resource Allocation for wireless Networks based on Distributed and Cooperative Relaying Units''.} 
Luca~Sanguinetti, \emph{Member, IEEE},
Marco~Luise, \emph{Fellow, IEEE},\thanks{G.~Bacci, L.~Sanguinetti and M.~Luise are with the Dip. Ingegneria dell'Informazione, University of Pisa, Via Caruso, 56126 Pisa, Italy (e-mail:
\{giacomo.bacci, luca.sanguinetti, marco.luise\}@iet.unipi.it). G.~Bacci is also with the Dept. Electrical Engineering, Princeton University, Olden Street, Princeton, NJ, 08544 USA (e-mail: gbacci@princeton.edu).} 
and H.~Vincent~Poor, \emph{Fellow, IEEE}\thanks{H.V. Poor is with the Dept. of Electrical Engineering, Princeton University, Olden Street, Princeton, NJ, 08544 USA (e-mail:
poor@princeton.edu).}\thanks{Part of this work has been submitted to the IEEE Wireless Communications and Networking Conference (WCNC), Shanghai, China, Apr. 2013.}}
\maketitle

\begin{abstract}
This work derives a distributed and iterative algorithm by which mobile terminals can selfishly control their transmit powers during the synchronization procedure specified by the IEEE 802.16m and the 3GPP-LTE standards for orthogonal frequency-division multiple-access technologies. The proposed solution aims at maximizing the energy efficiency of the network and is derived on the basis of a finite noncooperative game in which the players have discrete action sets of transmit powers. The set of Nash equilibria of the game is investigated, and a distributed power control algorithm is proposed to achieve synchronization in an energy-efficient manner under the assumption that the feedback from the base station is limited. Numerical results show that the proposed solution improves the energy efficiency as well as the timing estimation accuracy of the network compared to existing alternatives, while requiring a reasonable amount of information to be exchanged on the return channel.
\end{abstract}


\section{Introduction}\label{sec:intro}

The issue of energy efficiency has attracted a considerable interest in the information and telecommunication technology community during the last decade, as witnessed by the extensive literature available on this subject (see for example \cite{zhang10} and references therein). 
Among others, a challenge that lies in this paradigm is to prolong battery life of mobile terminals based on orthogonal frequency-division multiple-access (OFDMA) technologies such as those operating according to the IEEE 802.16m \cite{16m-2011} and the 3GPP long term evolution (LTE) \cite{3GPPTS36} standards. The first operation that must be accomplished by any terminal when joining the network is achieving correct synchronization with its serving base station (BS). This procedure is called initial ranging in IEEE 802.16m \cite{16m-2011}, and random access in LTE \cite{3GPPTS36}. It relies on a contention-based approach taking place over a specified set of subcarriers, which are used by each terminal to notify its entry request by transmitting a packet consisting of a randomly chosen code. Code identification as well as multiuser timing estimation are the main tasks of the BS during this procedure. These problems have received significant attention in the past few years, and some solutions are currently available in the literature (see for example \cite{fu07, ruan10, Sanguinetti12, sanguinetti12bis} and references therein). All the aforementioned works assume a deterministic increase of the transmit power upon successful synchronization without taking into account any energy efficiency issue. This is motivated by the fact that the energy efficiency problem in OFDMA-based technologies has been mainly analyzed for the data transmission phase (e.g., see \cite{miao11} and \cite{buzzi12} and references therein). A first attempt to reduce the power consumption during the initial synchronization phase can be found in \cite{BSLP12}, in which a low-complexity and iterative algorithm is proposed to allow each synchronization terminal (ST) and the BS to locally choose the transmit power and the detection strategy, respectively. The goal is to obtain a good tradeoff between detection capabilities and power consumption while satisfying quality-of-service (QoS) requirements given in terms of timing estimation error and probability of false code lock. The proposed solution is based on a noncooperative game-theoretic formulation and it is shown to provide significant gains in terms of reduced synchronization time and parameter estimation accuracy compared to existing alternatives based on a deterministic increase of the transmit power. Although interesting from a theoretical point of view, the analysis provided in \cite{BSLP12} is not suited for practical applications since it relies on the assumption of a continuous set of transmit powers. Moreover, comparisons with existing alternatives are carried out assuming that STs have perfect knowledge of the signal-to-interference-plus-noise ratio (SINR) measured at the BS. {A similar game-theoretic line of reasoning has been recently used for achieving synchronization in code-division multiple-access networks operating in a flat-fading scenario \cite{bacci12a}, and in a frequency-selective one \cite{bacci12c}.}

Motivated by the above considerations, in this work we return to the problem discussed in \cite{BSLP12} and extend both the power allocation approach and the numerical analysis as follows. We first assume that a \emph{finite} set of transmit powers is available at each terminal. Compared to \cite{BSLP12}, this more application-oriented assumption changes completely the nature of the energy-efficient optimization problem, as the tool of \emph{finite} noncooperative game theory is used to find its solution \cite{fudenberg91}. The set of Nash equilibria of the game is investigated and compared to that of the continuous-power noncooperative game discussed in \cite{BSLP12}. The theoretical analysis of the finite game is adopted to derive an iterative and distributed power allocation algorithm for achieving synchronization under the assumption of a \emph{limited feedback} from the BS. Numerical results are used to compare the performance of the proposed solution with that achieved by existing alternatives based on a deterministic increase of the transmit power (with and without contention resolution methods). It turns out that the proposed solution provides benefits in terms of energy efficiency and parameter estimation accuracy, using a reasonable amount of feedback resources.

The remainder of this paper is structured as follows.\footnote{The following notation is used throughout the paper. Matrices and vectors are denoted by boldface letters. ${\bf{I}}_n$, ${\bf{0}}_n$, and ${\bf{1}}_n$ are the $n\times n$ identity matrix, the $n\times1$ all-zero vector, and the $n\times1$ all-one vector, respectively, whereas $\mathbf{A}=\mathrm{diag}\{a(n)\,;\,\,n=1,2,\ldots ,N\}$ denotes an $N\times N$ diagonal matrix with entries $a(n)$ along its main diagonal. We use $\Exop\left\{\cdot \right\} $, $(\cdot )^{T}$ and $(\cdot )^{H}$ for expectation, transposition and Hermitian transposition, respectively, $\left\Vert \cdot \right\Vert $ for the Euclidean norm of the enclosed vector, $\left\lfloor x \right\rfloor$ to round $x$ to the nearest integer towards zero, $\left\lfloor x \right\rceil$ to round $x$ to the nearest integer, $\db[x]=10\log_{10} x$, and finally $[x]_a^b=\max(a,\min(x,b))$.} \sectionname~\ref{sec:ra} describes the system model and introduces the problem. Section III formulates the game and investigates its equilibria. The analysis is used in Section IV to derive an iterative and distributed synchronization algorithm whose performance assessment is provided in Section V. Finally, \sectionname~\ref{sec:conclusion} concludes the paper and discusses the applicability of this technique to current wireless standards.

\section{System model and problem formulation}\label{sec:ra}

\subsection{System model}\label{subsec:model}

We consider the uplink of an OFDMA-based system employing $N$ subcarriers with index set $\{0,1,\ldots,N-1\}$. To avoid aliasing problems, $2N_v$ null subcarriers are placed at the spectrum edges. The remaining $N-2N_v$ subcarriers are grouped into synchronization subcarriers and data subcarriers. The former are used by the STs entering the network through a \emph{contention-based synchronization procedure}, while the latter are assigned to mobile terminals for data transmission and channel estimation. We denote by $K$ the number of STs, and assume that the synchronization subcarriers are divided into $M$ subbands, each composed of a set of $V$ adjacent subcarriers, which is called a \emph{tile}. We denote by $\mathbf{c}_k=[c_k(0), \dots, c_k(MV-1)]$ the code chosen by the $k$th ST, and call $\theta_k$ the timing offset of the $k$th ST (normalized to the sampling period $T_s$).

As in \cite{BSLP12}, we consider a quasi-synchronous system in which no interblock interference (IBI) is present at the BS receiver, and we neglect any residual carrier frequency offset.\footnote{This assumption is reasonable as long as downlink estimation errors are within a few percents of the subcarrier spacing and low mobility applications are considered \cite{16m-2011}.} Moreover, we assume that the channel frequency response is nearly flat over each tile, and users other than those performing synchronization have been successfully synchronized to the BS so that they do not generate significant interference. Under the above assumptions, the vector $\mathbf{X}(m)$ containing the $m$th-tile discrete Fourier transform (DFT) outputs at the BS can be written as 
\begin{align}\label{eq:tileVector}
 \mathbf{X}(m) = \sum_{k=1}^K \sqrt{p_k} \mathbf{C}_k(m) \mathbf{a}(\theta_k) H_k(m) + \mathbf{n}(m)
\end{align}
where $p_k$ denotes the transmit power of the $k$th ST, $\mathbf{C}_k(m)=\diag\{ c_k(mV), \dots, c_k(mV+V-1) \}$, the vector $ \mathbf{a}(\theta_k)$ is given by
\begin{align}\label{eq:phaseVector}
 \mathbf{a}(\theta_k)=[1, e^{-j2\pi\theta_k/N},\dots,e^{-j2\pi(V-1)\theta_k/N}]^T
\end{align}
and $\mathbf{n}(m)$ is additive white Gaussian noise (AWGN) with zero mean and covariance matrix $\sigma_n^2 \mathbf{I}_V$. 



As mentioned above, the main tasks of the BS during the synchronization procedure are code detection and timing offset estimation. Following \cite{BSLP12}, the $k$th code $\mathbf{c}_k$ is declared as detected if the following generalized likelihood ratio test (GLRT) is satisfied:
\begin{align}\label{eq:glrt}
\frac{\Lambda_k(\hat{\theta}_k)}{\sum_{m=0}^{M-1}{\|\mathbf{X}(m)\|^2}}\ge\lambda
\end{align}
where the threshold $\lambda$ is a design parameter chosen so as to achieve a desired probability of false alarm $\overline{\Pi}_{\text{fa}}$, and $\Lambda (\hat{\theta}_k)$ is given by
\begin{align}\label{eq:Lambda}
\Lambda (\hat{\theta}_k)= \frac{1}{V}\sum_{m=0}^{M-1}{\left| \mathbf{a}^H( \hat{\theta}_k) \mathbf{C}^H_k(m) \mathbf{X}(m) \right|^2}.
\end{align}
In the above equation, $\hat{\theta}_k$ is the maximum likelihood estimate of $\theta_k$, given by \cite{BSLP12}
\begin{align}\label{eq:estphase}
 \hat{\theta}_k=\arg \max_{0\le \tilde \theta_k \le{\overline \theta}} \Lambda (\tilde {\theta}_k)
\end{align}
with $\overline\theta$ being the (normalized) round trip propagation delay for a user located at the cell boundary \cite{Morelli2004}.

\subsection{Problem formulation}\label{subsec:problem}

We let $\mathbf{p}=[p_1,p_2,\dots,p_K]^T$, and we define $\gamma_k$ as the SINR of the $k$th ST, given by \cite{BSLP12}
\begin{align}\label{eq:sinr} 
 \gamma_k=\nu_k(\mathbf{p}_{\setminus k})p_k
\end{align}
where $\mathbf{p}_{\setminus k}=\mathbf{p} \setminus p_k=[p_1,\dots,p_{k-1},p_{k+1},\dots,p_K]^T$, and $\nu_k(\mathbf{p}_{\setminus k})$ is defined as
 \begin{align}\label{eq:nu}
 \nu_k(\mathbf{p}_{\setminus k})=\frac{V\alpha_k}{\sigma_n^2+\sum_{\ell\neq k}{\alpha_\ell p_\ell}}
 \end{align}
with $
 \alpha_\ell={1}/{M}\sum\nolimits_{m=0}^{M-1}{|H_\ell(m)|^2}$
being the ST $\ell$'s average channel power gain across tiles.

Following \cite{BSLP12}, the energy-efficient optimization problem can be mathematically formalized for all STs $k\in\mathcal{K}=\{1,\dots,K\}$ as
\begin{align}\label{eq:problem}
 p_k^\star=\arg\max_{\substack{p_k\in\mathcal{P}_k}}\quad &\frac{\Pi_{\mathrm{d},k}(\gamma_k)}{p_k T}\\
 \label{eq:constraint}
 \textrm{subject to:}\quad&\mathsf{MSE}(\hat\theta_k)\le\overline{\mathsf{MSE}}_\theta\end{align}
where $\mathcal{P}_k$ denotes the set of transmit powers and $T$ is the duration of the cyclically extended OFDMA block, whereas $\mathsf{MSE}(\hat\theta_k)=\Exop\{|\hat\theta_k - \theta_k|^2\}$ is the mean-square error (MSE) of the timing estimate $\hat\theta_k$, and $\overline{\mathsf{MSE}}_\theta$ is the network QoS requirement in terms of maximum timing estimation MSE. In addition, $\Pi_{\mathrm{d},k}(\gamma_k)$ represents the probability of correct detection of code $\mathbf{c}_k$ given by \cite{BSLP12}
\begin{align} \label{eq:pd}
 \Pi_{d,k}\left(\gamma_k\right)&=I_{{\frac{(1+\gamma_k)(1-\lambda)}{1+(1-\lambda)\gamma_k}}}\left[M(V-1), M\right]
\end{align}
where $I_{x}[\cdot,\cdot]$ is the incomplete beta function \cite{abramowitz65}. Unlike \cite{BSLP12}, we assume that the power strategy set $\mathcal{P}_k$ is \emph{finite} and given by 
\begin{align}\label{eq:powerset}
 \mathcal{P}_k=\left\{ \pi_k^{(1)}, \pi_k^{(2)}, \dots, \pi_k^{(Q_k)} \right\}
\end{align}
where the number of power levels $Q_k$ is computed as
\begin{align}\label{eq:powerlevels}
 Q_k=1+\frac{\db[\overline p_k]-\db[\underline p_k]}{\db[\Delta_k]}
\end{align}
with $\Delta_k>1$ being the quantization step, and $\underline p_k$ and $\overline p_k$ the minimum and maximum power levels, respectively. From \eqref{eq:powerset} and \eqref{eq:powerlevels}, it follows that $\pi_k^{(1)}=\underline p_k$, $\pi_k^{(Q_k)}=\overline p_k$, and $\pi_k^{(q)}=\underline p_k\cdot\Delta_k^{q-1}$.

Setting $\mathcal{P}_k$ as specified in \eqref{eq:powerset} allows us to meet the technical requirements of practical systems in which transmit powers are usually equally spaced on a logarithmic scale to reduce the complexity of the front-end architecture and to increase the efficiency of power amplifiers (see for example the specifications provided by the IEEE 802.16m and 3GPP LTE standards in \cite{16m-2011} and \cite{3GPPTS36}, respectively). For simplicity, in all subsequent derivations we assume $\Delta_k=\Delta$, $\underline p_k=\underline p$ and $\overline p_k=\overline p$ for all $k\in\mathcal{K}$. This also implies $Q_k=Q$.


As mentioned previously, the aim of this work is to solve \eqref{eq:problem} taking into account the discrete nature of the power strategy sets $\{\mathcal{P}_k\}$. In the sequel, this is achieved by resorting to the analytical tools of \emph{finite} game theory \cite{fudenberg91}.

\section{Game formulation and analysis}\label{subsec:ne}

Using the results illustrated in \cite{BSLP12}, it follows that the MSE in \eqref{eq:problem} can be met provided that 
 \begin{align}\label{eq:powermin}
 p_k\ge\frac{\gamma_{\mathrm{req}}}{\nu_k(\mathbf{p}_{\setminus k})}
 \end{align}
 where $\gamma_{\mathrm{req}}$ denotes the minimum SINR such that the MSE constraint \eqref{eq:constraint} is satisfied with equality, i.e., $\mathsf{MSE}(\hat\theta_k)=\overline{\mathsf{MSE}}_\theta$. In particular, $\gamma_{\mathrm{req}}$ is found to be \cite{BSLP12}
 \begin{align}\label{eq:gammamin}
 \gamma_{\mathrm{req}} = \frac{3N^2}{2M\pi^2(V^2-1)} \cdot \frac{1}{\rho}
 \end{align}
 where $ \rho$ is defined as $\rho=\overline{\mathsf{MSE}}_\theta-\mu^2(\hat\theta_k)$
 with $\mu(\hat\theta_k)=\Exop\{\hat\theta_k\}-\theta_k$ denoting the bias of the timing estimate $\hat\theta_k$. Using \eqref{eq:powermin}, the optimization problem in \eqref{eq:problem} can be reformulated as
\begin{align}\label{eq:problem2}
 p_k^\star=\arg\max_{p_k\in{\mathcal{A}}_k(\mathbf{p}_{\setminus k})}\quad \frac{\Pi_{\mathrm{d},k}(\gamma_k)}{p_kT}
\end{align}
where
\begin{align}\label{eq:powersubset}
 {\mathcal{A}}_k(\mathbf{p}_{\setminus k})=\left\{p_k\in\mathcal{P}_k: p_k\ge\frac{\gamma_{\mathrm{req}}}{\nu_k(\mathbf{p}_{\setminus k})}\right\}
\end{align}
is the power strategy subset that allows the $k$th ST to meet the MSE constraint in \eqref{eq:problem}. For notational simplicity, in all subsequent derivations we omit the functional dependence of ${\mathcal{A}}_k$ on $\mathbf{p}_{\setminus k}$.

The power allocation problem in \eqref{eq:problem2} can be formulated as a generalized\footnote{The game is generalized since ${\mathcal{A}}_k$ depends on the other STs' power $\mathbf{p}_{\setminus k}$ (see for example\cite{fudenberg91} and \cite{facchinei07} for more details).} noncooperative game with complete information \cite{fudenberg91}, denoted by $\mathcal{G}=[\mathcal{K}, \{\mathcal{A}_k\}, \{u_k\}]$. In particular, $\mathcal{K}=\{1,2, \dots,K\}$ is the player set, $\mathcal{A}_k$ is the action set of the $k$th player, and $u_k$ is its payoff function given by
 \begin{align}\label{eq:utility}
 u_k(\mathbf{p})=\frac{\Pi_{\mathrm{d},k}(\gamma_k)}{p_k T}
 \end{align}
which depends on the power allocation $\mathbf{p}$ through $\gamma_k$ as in \eqref{eq:sinr}. The discrete nature of $\mathcal{A}_k$ places $\mathcal{G}$ into the category of \emph{finite} generalized noncooperative games. As mentioned previously, this is much different from \cite{BSLP12}, in which the \emph{continuous} nature of the action sets allows us to formalize the optimization problem as an \emph{infinite} generalized noncooperative game $\mathcal{G}_\mathrm{c}$.

\subsection{Analysis of the equilibria}\label{subsec:analysis_ne}

The existence and uniqueness of the generalized Nash equilibria (GNE) of $\mathcal{G}$ are studied in the following. The analysis is conducted only for \emph{pure} (i.e., deterministic) strategies. This choice is motivated by the fact that in compact strategy spaces, \emph{mixed} (i.e., statistical) strategies are generally less attractive due to implementation difficulties in wireless communications systems \cite{lasaulce11}. 

To proceed further, we recall that a vector $\mathbf{p}^\star=\left[p_1^\star,p_2^\star,\dots,p_K^\star\right]^T$ is a pure-strategy GNE of $\mathcal{G}$ if, for any $k\in\mathcal{K}$,
 \begin{equation}\label{eq:gne} 
 u_k\left([p_k^\star,\mathbf{p}_{\setminus k}^\star]\right) \ge u_k\left([p_k,\mathbf{p}_{\setminus k}^\star] \right)
\end{equation}
for all transmit powers $p_k\in \mathcal{A}_k$. Another way to define a pure-strategy GNE is to make use of the concept of \emph{best response} \cite{fudenberg91}. In particular, we have that a vector $\mathbf{p}^\star$ is a GNE if each element $p_k^\star$ is the best response $r_k(\mathbf{p}_{\setminus k}^\star)$ to the powers $\mathbf{p}_{\setminus k}^\star$ chosen by the other players, with $r_k(\mathbf{p}_{\setminus k}^\star)$ being the solution of the following problem:
\begin{align}
 \label{eq:br}
r_k(\mathbf{p}_{\setminus k}^\star) = \arg\max_{\tilde p_k\in{\mathcal{A}}_k} \frac{\Pi_{\textrm{d},k}(\nu_k(\mathbf{p}^\star_{\setminus k})\tilde p_k)}{\tilde p_kT}
\end{align}
in which we have used \eqref{eq:utility}, and we have explicitly written the functional dependence of the detection probability $\Pi_{\textrm{d},k}$ on $\tilde p_k$ and $\mathbf{p}_{\setminus k}^\star$ through $\gamma_k$ in \eqref{eq:sinr}.

\begin{theorem}\label{th:existence}
 Let us define the SINR $\gamma^\star$ as
 \begin{align}\label{eq:gammaStar}
 \gamma^\star=\max(\gamma_{\mathrm{req}},\tilde\gamma)
 \end{align}
 with $\tilde\gamma$ being the solution of
 \begin{align}\label{eq:gammaTilde}
 \left.\frac{\partial \Pi_{\textrm{d},k}(\gamma)}{\partial\gamma}\right|_{\gamma=\tilde\gamma}=\frac{\Pi_{\textrm{d},k}(\tilde\gamma)}{\tilde\gamma}.
 \end{align}
 Then, the game $\mathcal{G}$ admits pure-strategy GNE provided that
 \begin{align}\label{eq:size}
 \gamma^\star(K-1)<V.
 \end{align}
  The proof can be found in \appendixname~\ref{app:existence}.\hfill$\blacksquare$
\end{theorem}

\begin{theorem}\label{th:uniqueness}
Let ${\mbox{$\mathcal{E}$}}^\star$ be the set of pure GNE for $\mathcal{G}$. Then, the cardinality of ${\mbox{$\mathcal{E}$}}^\star$ is such that
 \begin{align}\label{eq:uniqueness}
 |{\mbox{$\mathcal{E}$}}^\star|\ge1.
 \end{align}
  The proof can be found in \appendixname~\ref{app:uniqueness}.\hfill$\blacksquare$
\end{theorem}

\theoremname~\ref{th:existence} provides a \emph{sufficient} condition for the set of GNE not to be empty, and \theoremname~\ref{th:uniqueness} states that the GNE is not necessarily unique. This means that the uniqueness property proven in \cite{BSLP12} for $\mathcal{G}_\mathrm{c}$ no longer holds for $\mathcal{G}$. In other words, quantizing the set of actions makes the game $\mathcal{G}$ lose the uniqueness property for the GNE. In this context, it is interesting to show the following result.

\begin{theorem}\label{th:PoS}
 Among all $\mathbf{p}^\star\in{\mbox{$\mathcal{E}$}^\star}$, the smallest component-wise GNE $\mathbf{p}^\star_\Delta$ is such that
  \begin{align}\label{eq:so}
    \mathbf{p}_\Delta^\star=\arg\max_{\mathbf{p}^\star\in{\mbox{$\mathcal{E}$}}^\star}\,\,\sum_{k=1}^{K}{u_k(\mathbf{p}^\star)}.
  \end{align}
The proof can be found in \appendixname~\ref{app:PoS}.\hfill$\blacksquare$
\end{theorem}
The above result states that $\mathbf{p}_\Delta^\star$ is the best GNE in terms of social welfare (joint optimization) or, equivalently, it is the most efficient GNE in a social sense \cite{lasaulce09}. Note that this does not amount to saying that $\mathbf{p}_\Delta^\star$ is the socially optimum solution of \eqref{eq:problem}, as noncooperative equilibria are known to be generally inefficient \cite{lasaulce09}. Improving the equilibrium efficiency is out of the scope of this paper and is left as a future work.

\subsection{Numerical analysis}\label{numerical_analysis_equilibria}

Unlike the unique GNE of $\mathcal{G}_\mathrm{c}$ in \cite{BSLP12}, the multiple equilibria of $\mathcal{G}$ cannot be expressed in a closed form as a function of the network parameters because of the $\argmax$ operator in \eqref{eq:br}. A numerical analysis is thus conducted to make comparisons and to evaluate the impact of the discretization of the action sets. To this aim, we concentrate on the optimal (in a social sense) $\mathbf{p}_\Delta^\star$ and resort to the exhaustive search method described in \cite{young04} to solve \eqref{eq:so}. The numerical results are averaged over $20,000$ independent realizations of a network with the following parameters: $T_s=89.28\ns$, $N=1024$, $M=4$, $V=36$, $\overline\Pi_{\textrm{fa}}=10^{-5}$, and $\overline{\mathsf{MSE}}_\theta=324$, which yield $\rho=128$, $\lambda=0.12$, $\db[\gamma_{\mathrm{req}}]=-6.19$, and $\db[\gamma^\star]=\db[\tilde\gamma]=7.09$ (see \cite{BSLP12} for a detail discussion on this parameter setting). The normalized power constraints are fixed to $\db[\underline p/\sigma_n^2]=-20$ and $\db[\overline p/\sigma_n^2]=+30$ for all $k$, whereas the ST distances $d_k$ are randomly chosen from a uniform distribution in $[R/10, R]$ with $R$ being the cell radius. The channel power gains are normalized to a distance $R/2$ and are modeled using the $6$-tap ITU modified vehicular-A model \cite{ITU97} with a path loss exponent $\varsigma=2$.


\figurename~\ref{fig:nmse} reports the normalized MSE defined as $\textsf{NMSE}(\mathbf{p}_\mathrm{c}^\star)=\Exop\{\|\mathbf{p}_\mathrm{c}^\star-\mathbf{p}_\Delta^\star\|^2/\|\mathbf{p}_\mathrm{c}^\star\|^2\}$ as a function of $K$ for different quantization steps, where $\mathbf{p}_\mathrm{c}^\star$ is the unique GNE of $\mathcal{G}_\mathrm{c}$. The maximum number of STs is fixed to $\lfloor 1+V/\gamma^\star \rfloor=8$. Observe that the condition $K \le 8$ is required to meet \eqref{eq:size}. In addition, it represents a necessary and sufficient condition for the existence of the unique GNE $\mathbf{p}_\mathrm{c}^\star$ \cite{BSLP12}. As expected, $\textsf{NMSE}(\mathbf{p}_\mathrm{c}^\star)$ decreases as $\Delta$ becomes smaller since the discrete action sets $\mathcal{A}_k$ in $\mathcal{G}$ tend to better approximate the continuous ones in $\mathcal{G}_\mathrm{c}$. As can be seen, $\textsf{NMSE}(\mathbf{p}_\mathrm{c}^\star)$ increases as $K$ increases, meaning that the difference between $\|\mathbf{p}_\Delta^\star\|^2$ and $\|\mathbf{p}_\mathrm{c}^\star\|^2$ becomes larger as the number of STs increases. In particular, we see that $\textsf{NMSE}(\mathbf{p}_\mathrm{c}^\star)$ is almost constant up to $K=4$, whereas it rapidly increases for larger values. 

To evaluate the impact of this difference on the system performance in terms of social welfare, \figurename~\ref{fig:utility} reports the experimental $\sum_{k}{u_k(\mathbf{p}_\Delta^\star)}/\sum_{k}{u_k(\mathbf{p}_\mathrm{c}^\star)}$ as a function of $K$ in the same operating conditions of \figurename~\ref{fig:nmse}. As seen, the ratio $\sum_{k}{u_k(\mathbf{p}_\Delta^\star)}/\sum_{k}{u_k(\mathbf{p}_\mathrm{c}^\star)}$ is approximately $1$ for $K\le 6$ while it increases for larger values of $K$. A similar behavior (not shown for the sake of brevity) is observed if the user-basis ratio ${u_k(\mathbf{p}_\Delta^\star)}/{u_k(\mathbf{p}_\mathrm{c}^\star)}$ is considered. From these results, it follows that limiting the STs to use a discrete set of strategies $\mathcal{A}_k$ increases the global system performance rather than introducing a detrimental effect. This phenomenon is known as a Braess-type paradox \cite{perlaza11}, and it has already been observed in other different contexts (such as routing in \cite{roughgarden05, altman06} and wireless communications in \cite{altman08, rose11bis}). Roughly speaking, the Braess-type paradox occurs because the average number of GNEs increases when $K$ becomes larger. For example, when $\db[\Delta]=1$, the numerical results indicate that the average number of GNE for $K=\{2,3,4,5,6,7, 8\}$ is $\{1.0, 1.1, 1.2, 1.3, 1.7, 2.7, 20.9\}$, respectively.

\section{Energy-efficient distributed synchronization}\label{sec:algo}

We now show how to exploit the GNE analysis provided so far to derive a practical power control algorithm for achieving synchronization in the network modeled as in \sectionname~\ref{subsec:model}. We start denoting by $p_k[n]$ the transmit power of the $k$th ST at the $n$th iteration step. Then, we observe that, using the results illustrated in \cite{yates95} and \cite{altman03}, it can be easily proven that, under hypothesis \eqref{eq:size}, an iterative algorithm operating according to the best-response dynamic in \eqref{eq:br}, i.e.,  
\begin{align}\label{eq:br5}
     p_k[n+1]=\arg\max_{\tilde p_k\in\mathcal{A}_k} \frac{\Pi_{\mathrm{d}}\left(  \nu_k(\mathbf{p}_{\setminus k}[n])\tilde p_k\right)}{\tilde p_kT}
\end{align}
converges to the most socially efficient GNE $\mathbf{p}_\Delta^\star$ if ${p}_k[0]=\underline{p}$ for $k=1,2,\ldots, K$. 
Note that the computation of $p_k[n+1]$ in \eqref{eq:br5} requires knowledge of $\nu_k(\mathbf{p}_{\setminus k}[n])$. Using \eqref{eq:sinr}, it follows that $\nu_k(\mathbf{p}_{\setminus k}[n])$ can be obtained as 
\begin{align}\label{eq:br3}
  \nu_k(\mathbf{p}_{\setminus k}[n])=\frac{\gamma_k[n]}{p_k[n]}
\end{align}
where $\gamma_k[n]$ is ST $k$'s SINR measured at the BS at time step $n$. While $p_k[n]$ is locally available at the transmitter, $\gamma_k[n]$ can only be estimated at the BS and sent to the $k$th ST on a downlink control channel. Following \cite{BSLP12}, an unbiased estimate $ \hat{\gamma}_k[n]$ of $\gamma_k[n]$ can be computed as
\begin{align}\label{eq:estsinr}
 \hat{\gamma}_k[n]=\frac{V\Lambda_k(\hat{\theta}_k)-\sum_{m=0}^{M-1}{\|\mathbf{X}(m)\|^2}}{\sum_{m=0}^{M-1}{\|\mathbf{X}(m)\|}^2-\Lambda_k(\hat{\theta}_k)}.
\end{align}
To reduce the amount of information to be exchanged, we assume the quantity $ \hat{\gamma}_k[n]$ to be quantized on a logarithmic scale using a uniform $B$-bit quantizer. This produces
\begin{align}\label{eq:estsinr_quantized}
\db[{\mu_k[n]}] = \db[\Delta_\gamma] \cdot b_k[n]
\end{align} 
where
\begin{align}
\label{eq:b_k[n]}
b_k[n]=\left\lfloor \frac{\left.{\left( \left[\hat{\gamma}_k[n]\right]_{\underline\gamma}^{\overline\gamma} - \underline{\gamma} \right)}\right|_{\textrm{dB}}}{\db[\Delta_\gamma]} \right\rceil
\end{align}
and
\begin{align}
\label{eq:gamma_res}
\db[\Delta_\gamma] = \frac{\db[{\overline\gamma}] - \db[{\underline\gamma}]}{2^{B} -1}
\end{align} 
is the quantizer resolution (also known at the ST side, e.g., selected by the system standard), with $\overline\gamma$ and  $\underline\gamma$ being the maximum and minimum expected values for $\hat\gamma_k[n]$, respectively. The BS sends on a broadcast downlink channel the integer $b_k[n]$, which is used by the $k$th ST to retrieve the quantized version of $\hat\gamma_k[n]$ using \eqref{eq:estsinr_quantized}. 

Replacing $\gamma_k[n]$ with $\mu_k[n]$ into \eqref{eq:br3} and substituting the result into \eqref{eq:br5}, we eventually obtain
\begin{align}\label{eq:br6}
     p_k[n+1]=\arg\max_{\tilde p_k\in\mathcal{A}_k} \frac{\Pi_{\mathrm{d}}\left(\mu_k[n] \tilde p_k / {p_k[n]}\right)}{\tilde p_kT}
\end{align}
Recalling \eqref{eq:estsinr_quantized}, it follows that its evaluation at the ST side requires only the knowledge of $b_k[n]$.

\begin{algorithm}[t]
\caption{Discrete and limited feedback best response synchronization algorithm (DLF-BRSA)}
\begin{enumerate}
\item \emph{Initialization:} each ST $k=1,2,\ldots,K$
 \begin{enumerate}
 \item initializes the transmit power $p_k[0]$ to the lowest power value $\underline p$;
 \item sets $n=0$.
 \end{enumerate}
 \item \emph{Iterative algorithm:} at each step $n$, each ST $k$
 \begin{enumerate}
 \item receives on a common downlink channel the result of the GLRT \eqref{eq:glrt} for code $\mathbf{c}_k$ and the integer $b_k[n]$ computed through \eqref{eq:estsinr} -- \eqref{eq:gamma_res};
 \item if the GLRT for code $\mathbf{c}_k$ is verified and $\mu_k[n]>\gamma_{\mathrm{req}}$, exits the algorithm (i.e., ST $k$ is successfully associated to the BS), otherwise goes to the next step;
 \item adjusts its transmit power according to \eqref{eq:br6};
 \item updates $n=n+1$.
\end{enumerate}
\end{enumerate}
\label{alg:brsa}
\end{algorithm}
 
Collecting all the above facts together leads to the energy-efficient synchronization algorithm illustrated in Algorithm \ref{alg:brsa}, which allows each ST to operate in a complete distributed manner without any knowledge of other users' power allocation strategies (as if in a single-user scenario). 
Observe that Algorithm \ref{alg:brsa} is reminiscent of the best-response synchronization algorithm (BRSA) illustrated in \cite{BSLP12}, except for the discrete action sets and the limited feedback from the BS, which makes it more suited for a practical implementation. In the sequel, we call the iterative procedure described in Algorithm \ref{alg:brsa} as discrete and limited feedback best response synchronization algorithm (DLF-BRSA).

\begin{remark}\label{rem1} It is worth observing that, in their most basic forms, iterative algorithms based on best-response dynamics require a significant amount of information to be locally available at the player (transmitter) \cite{rose11}. For example, they usually require knowledge of the number of players and of the actions played by all the other players. To overcome this problem, other algorithms based on reinforcement learning techniques have been adopted in the literature \cite{rose11}. The main advantage of these solutions is that they do require each player to know only its corresponding utility. Although based on best-response dynamics, DLF-BRSA possesses most of the advantages shown by other reinforcement learning-based algorithms, as it allows each ST to operate in a distributed and iterative way requiring only knowledge of its own estimated SINR, as illustrated in Algorithm \ref{alg:brsa}.
\end{remark}


\begin{remark} Most of the computational complexity of DLF-BRSA is represented by the exhaustive search in \eqref{eq:br6}, which must be performed at each iteration step $n$ over the $Q$ discrete power levels of the set $\mathcal{A}_k$. In those applications characterized by large values of $Q$, this may represent an implementation impairment. In such cases, one may resort to the supermodularity properties of the utility function $u_k$ (see \appendicesname~\ref{app:existence} and \ref{app:uniqueness}) and reduce the search complexity looking only at the values of $\tilde p_k\in {\mathcal{A}}_k$ in the neighborhood of $p_k[n]$.
\end{remark}

\section{Simulation results}\label{sec:results}
Numerical simulations are now used to assess the performance of DLF-BRSA and to make comparisons with existing alternatives. As in \sectionname~\ref{numerical_analysis_equilibria}, the numerical analysis is conducted by averaging over $20,000$ independent realizations of a network whose parameters are fixed as follows: $T_s=89.28\ns$, $N=1024$, $M=4$, $V=36$, $\overline\Pi_{\textrm{fa}}=10^{-5}$, and $\overline{\mathsf{MSE}}_\theta=324$, which yield $\rho=128$, $\lambda=0.12$, $\db[\gamma_{\mathrm{req}}]=-6.19$, and $\db[\gamma^\star]=\db[\tilde\gamma]=7.09$. {The minimum and maximum expected values of $\hat \gamma_k[n]$ in \eqref{eq:gamma_res}, based on an extensive simulation campaign, are fixed to $\db[{\underline \gamma}] = -8$ and $\db[{\overline \gamma}] = +16$, whereas the appropriate number of bits $B$ is chosen later on the basis of the following numerical analysis.} Without loss of generality, we concentrate on the first ST (i.e., $k=1$) and assess the performance of the investigated solutions when its distance $d_1$ is kept constant. All other STs are assumed to be randomly located in $[R/10, R]$ with $R$ being the cell radius. The normalized power constraints are fixed to $\db[\underline p/\sigma_n^2]=-20$ and $\db[\overline p/\sigma_n^2]=+30$, and the same power initialization $p_k[0]=\underline p$ is used for all STs $k\in\mathcal{K}$, which also use a common power quantization step $\db[\Delta]=1$.


{\figurename~\ref{fig:powerVsUser} reports the average normalized power expenditure $p_{\textrm{avg}}/\sigma^2_n$ (in dB) required by DLF-BRSA for successfully completing the synchronization procedure. The numerical results are plotted as functions of $K$ for $B = \{1, 2, 3, 8\}$. The results obtained with DLF-BRSA when $B\rightarrow\infty$ (i.e., with continuous-SINR feedback) are used as a benchmark. Comparisons are also made with the BRSA illustrated in \cite{BSLP12} in which the action sets are continuous and perfect knowledge of the estimated SINRs is available at the STs. The results of {\figurename~\ref{fig:powerVsUser} indicate that the quantization of the SINRs has only a marginal effect on the performance of DLF-BRSA. In fact, it has practically the same performance for $B = 3,8$ and $B \rightarrow\infty$, whereas a significant degradation is observed only for $B = 1$. We argue that the quantization of the estimated SINRs marginally impacts the system performance since it is basically perceived at the STs as an additional estimation error introduced by the BS (which on the other hand can actually exploit real-valued estimation methods). Based on the above results, in all subsequent simulations we set $B=3$. From \eqref{eq:gamma_res}, recalling that $\db[{\underline \gamma}] = -8$ and $\db[{\overline \gamma}] = +16$, we have $\db[\Delta_\gamma] = 3.43$.

{To evaluate the impact of the discretization of the action sets, we now compare the performance of BRSA with those of DLF-BRSA. From \figurename~\ref{fig:powerVsUser}, it follows that they do perform identically when the DLF-BRSA uses $B\ge3$. This seems to contradict the numerical results of \figurename~\ref{fig:utility}, discussed at the end of \sectionname~\ref{numerical_analysis_equilibria}, which show that discretizing the set of strategies is beneficial for individual (and, consequently, global) performance. On the basis of the analysis of \sectionname~\ref{numerical_analysis_equilibria}, the DLF-BRSA is expected to outperform the BRSA. The motivation behind this contradictory result can be understood by recalling that the considered ST takes part in the synchronization procedure as long as it is not correctly detected by the BS. As a consequence, what really impacts on the performance of DLF-BRSA and BRSA is its power evolution from the time it enters the network ($n=0$) to the time step $n^\textrm{exit}$ in which the exit conditions (detailed in Step b2 of DLF-BRSA) are satisfied. 

To this aim, \figurename~\ref{fig:timeVsUser} reports the average number of iterations $n^{\textrm{exit}}_{\textrm{avg}}$ as a function of $K$, which turns out to be the same for the both the BRSA and the DLF-BRSA (with $B\ge3$) with good approximation. Interestingly, numerical simulations confirm that, when $0 \le n \le n^\textrm{exit}$, the difference between the power updates across the two schemes is negligible. On the contrary, the performance measured in \figurename~\ref{fig:utility} corresponds to that achieved by the BRSA and the DLF-BRSA schemes \emph{without} the exit conditions listed in Step b2 (as the GNE, computed in \sectionname~\ref{numerical_analysis_equilibria} through an exhaustive search \cite{young04}, can also be achieved using the best-response dynamics described in \sectionname~\ref{sec:algo}), whose convergence time is usually much higher than $n^\textrm{exit}$. This is the reason why the performance in terms of total energy expenditure reported in \figurename~\ref{fig:powerVsUser} is similar in the two cases. On the basis of the above results, we can conclude that DFL-BRSA yields the same performance of BRSA. However, this is achieved while \emph{i})~reducing the complexity of the user terminals (thanks to the discretization of the power amplifier), and \emph{ii})~requiring a limited amount of feedback from the BS (thanks to the finite number of bits $B$ used to send the estimated SINRs).

The performance of DLF-BRSA are now compared with those achieved by two alternative solutions based on a deterministic increase of the transmit power: the deterministic synchronization algorithm (DSA), in which the update rule is $\db[{p_k[n+1]}]=\db[{p_k[n]}]+\db[{\Delta}]$; and the binary exponential backoff DSA (BEB-DSA), in which $\db[{p_k[n+n_e]}]=\db[{p_k[n]}]+\db[{\Delta}]$, where $n_e$ is an exponentially-distributed backoff counter (see \cite{BSLP12} for more details). In all subsequent simulations, we assume $K=5$ and set $\db[\Delta]=1$ and $p_k[0]=\underline p$ for $k=1,2,\ldots,K$.



\figurename~\ref{fig:powerVsDistance} shows $p_{\textrm{avg}}/\sigma_n^2$ for all investigated solutions as a function of the normalized distance $d_1/R$, and \figurename~\ref{fig:timeVsDistance} illustrates the average time $T_{\textrm{avg}}$ needed to complete the synchronization procedure. In particular, $T_{\textrm{avg}}$ is computed as $T_{\textrm{avg}}= T_f\cdot n^{\textrm{exit}}_{\textrm{avg}}$, where $T_f = 5$ ms accounts for the time interval (frame time) between two successive synchronization attempts. In addition, \figurename~\ref{fig:mseVsDistance} shows the MSE of the timing estimate $\hat\theta_1$ for different values of $d_1/R$. 

From the results of \figurename~\ref{fig:powerVsDistance}, it follows that DLF-BRSA provides roughly the same power consumption of BEB-DSA, which is significantly lower than that needed by DSA. However, the results of \figurename~\ref{fig:timeVsDistance} show that the time required by DLF-BRSA to achieve synchronization is much shorter than that needed by DSA and BEB-DSA, especially when $d_1/R$ increases. In addition, \figurename~\ref{fig:mseVsDistance} shows that the estimation accuracy with DLF-BRSA is higher than that with both DSA and BEB-DSA. Collecting all the above facts together, we may conclude that DLF-BRSA provides better results in terms of \emph{energy efficiency} and \emph{parameter estimation accuracy}, also providing some performance that slightly depends on the transmitter-receiver distance. This is achieved at the price of a slight increase of information to be fed back over the control channel. In particular, the amount of information to be exchanged during each frame and for each $\mathbf{c}_\ell\in\mathcal{C}$ is the following: $1$ bit to broadcast the outcome of the GLRT, and $B=3$ bits to transmit the quantized SINRs. This means that a total of $4|\mathcal{C}|$ bits per frame time $T_f$ is required by DLF-BRSA, which corresponds to a feedback rate on the order of a few tens of kb/s, given that $|\mathcal{C}|$ is usually on the order of tens to hundreds (e.g., see \cite{16m-2011, 3GPPTS36}).

\section{Conclusion}\label{sec:conclusion}

In this work, we have formalized the power allocation problem for energy-efficient contention-based synchronization in OFDMA-based networks as a finite constrained noncooperative game. The generalized Nash equilibria have been analytically studied, and numerically evaluated. The above results have been used to derive a distributed and iterative energy-efficient power control algorithm with discrete powers and limited feedback. The performance of the above solution have been evaluated and compared with alternatives by means of numerical simulations. Using realistic system parameters and widely agreed-upon channel models, we have shown that the proposed solution incurs only a negligible degradation with respect to the scheme illustrated in \cite{BSLP12}, while a significant gain is achieved with respect to deterministic-based power allocation approaches (both with and without contention resolution methods). The derived technique requires a feedback on the downlink on the order of a few tens of kb/s, which can be easily accommodated in current IEEE 802.16m \cite{16m-2011} and LTE \cite{3GPPTS36} standards. Since the proposed solution shows a (much) faster synchronization time than deterministic methods, it can be used to further increase the energy efficiency of mobile terminals by reducing the frequency of periodic ranging procedures, that are currently used by 4G communication systems to meet the transmission latency requirements.

\appendices

\section{Proof of \theoremname~\ref{th:existence}}\label{app:existence}

\figurename~\ref{fig:existence} shows a pictorial representation of the typical shape of the utility $u_k(\mathbf{p})$ as a function of the power $p_k=\pi_k^{(q)}$ for a fixed interference $\mathbf{p}_{\setminus k}$ (all quantities are on a logarithmic scale, although the subscript `dB' is suppressed for the sake of presentation). The relevant points of the utility function in terms of SINR, scaled by the quantity $\nu_k(\mathbf{p}_{\setminus k})$, are also reported: in addition to $\gamma_{\mathrm{req}}$ and $\tilde\gamma$, defined in \eqref{eq:gammamin} and \eqref{eq:gammaTilde}, respectively, \figurename~\ref{fig:existence} also shows the inflection point $\dot\gamma$, $\dot\gamma<\tilde\gamma\le\gamma^\star$, such that $\Pi_{\textrm{d},k}(\gamma_k)$ is strictly convex for $\gamma_k<\dot\gamma$, and strictly concave for $\gamma_k>\dot\gamma$. Although \figurename~\ref{fig:existence} depicts the case $\gamma_{\mathrm{req}}<\tilde\gamma$, the considerations drawn in the following apply in the case $\gamma_{\mathrm{req}}\ge\tilde\gamma$ as well. Circular markers report $u_k([\pi_k^{(q)}, \mathbf{p}_{\setminus k}])$ for $q=1,\dots, Q$ (in this example, $Q=8$). Note that the best-response map defined in \eqref{eq:br} may yield $r_k(\mathbf{p}_{\setminus k})<\gamma^\star/\nu_k(\mathbf{p}_{\setminus k})$, as occurs in this example.

A GNE in the game $\mathcal{G}$ exists provided that the $K$ sets ${\mathcal{A}}_k(\mathbf{p}_{\setminus k})\subseteq{\mathcal{P}}_k$, $k\in\mathcal{K}$, are nonempty, which translates, using \eqref{eq:powersubset}, into ensuring that there exists at least a power level $\pi_k^{(q)}\in{\mathcal{P}}_k$ such that $\pi_k^{(q)}\ge\gamma_{\mathrm{req}}/\nu_k(\mathbf{p}_{\setminus k})$ for all $k$. Since $\gamma_{\mathrm{req}}\le\gamma^\star$ by hypothesis, it is sufficient to show that $\gamma^\star/\nu_k(\mathbf{p}^\star_{\setminus k})\le r_k(\mathbf{p}^\star_{\setminus k})=p_k^\star$ for all $k\in\mathcal{K}$. By following the same steps as in \cite{bacci12a}, we can derive the sufficient condition \eqref{eq:size}, which becomes also necessary in the case $\gamma^\star=\gamma_{\mathrm{req}}$. Note that, unlike \cite{bacci12a}, here we cannot derive a necessary condition that holds for any $\tilde\gamma>\gamma_{\mathrm{req}}$, because of the inequality $\gamma^\star/\nu_k(\mathbf{p}_{\setminus k})\le r_k(\mathbf{p}_{\setminus k})$ that is originated from using a finite set, and hence GNE might exist even though \eqref{eq:size} is not fulfilled. This is also in accordance to what highlighted in \eqref{eq:br}, as an equilibrium can exist even if $p_k^\star<\gamma^\star/\nu_k(\mathbf{p}^\star_{\setminus k})$, provided that $\gamma^\star>\gamma_{\mathrm{req}}$.

To proceed further with the proof of existence, it is useful to introduce the following definition:
\begin{definition}[\!\!\cite{altman03}]\label{def:ascending}
 A best response $r_k(\mathbf{p}_{\setminus k})$ possesses the \emph{ascending property} if $r_k(\mathbf{p}_{\setminus k})\le r_k(\mathbf{p}^\prime_{\setminus k})$ for all $k\in\mathcal{K}$ when $\mathbf{p}_{\setminus k} \neq \mathbf{p}^\prime_{\setminus k}$ is such that $p_\ell\le p^\prime_\ell$ $\forall \ell\neq k$.\hfill$\blacksquare$
\end{definition}

To show that the best response \eqref{eq:br} is ascending, let us define $r_k=r_k(\mathbf{p}_{\setminus k})$, $r_k^\prime=r_k(\mathbf{p}^\prime_{\setminus k})$, $\nu_k=\nu_k(\mathbf{p}_{\setminus k})$, and $\nu^\prime_k=\nu_k(\mathbf{p}^\prime_{\setminus k})$ for notational convenience. Note that assuming $\mathbf{p}_{\setminus k} \neq \mathbf{p}^\prime_{\setminus k}$ with $p_\ell\le p^\prime_\ell$ $\forall \ell\neq k$ implies $\nu_k>\nu^\prime_k$. Let us consider two different cases:
\begin{enumerate}
\item $r_k<\gamma^\star/\nu_k$: in this domain, $u_k([p_k, \mathbf{p}^\prime_{\setminus k}])$ is an increasing function of $p_k$. Hence, power vector $\mathbf{p}^\prime_{\setminus k}$'s best response is $r_k^\prime=\arg\max_{p_k} u_k([p_k, \mathbf{p}^\prime_{\setminus k}])\ge r_k$.
\item $r_k\ge\gamma^\star/\nu_k$: in this domain, since $\nu_k>\nu^\prime_k$, both $u_k([p_k, \mathbf{p}_{\setminus k}])$ and $u_k([p_k, \mathbf{p}^\prime_{\setminus k}])$ are decreasing functions of $p_k$. Hence, for $r_k$ to be a best response, the condition $u_k([r_k, \mathbf{p}_{\setminus k}])>u_k([r_k/\Delta, \mathbf{p}_{\setminus k}])$ must hold, with $r_k/\Delta<\gamma^\star/\nu_k$. Due to the asymmetry of $u_k([p_k, \mathbf{p}_{\setminus k}])$ with respect to the point of maximum $\gamma^\star/\nu_k$, $r_k/\Delta<\check{p}_k$, where $\check{p}_k$, $2\gamma^\star/\nu_k-r_k<\check{p}_k<\gamma^\star/\nu_k<\gamma^\star/\nu^\prime_k$, is the power level such that $u_k([\check{p}_k, \mathbf{p}_{\setminus k}])=u_k([r_k, \mathbf{p}_{\setminus k}])$. Since $u_k([p_k, \mathbf{p}^\prime_{\setminus k}])$ is increasing in the region $p_k<\gamma^\star/\nu^\prime_k$, $u_k([\check{p}_k, \mathbf{p}_{\setminus k}])<u_k([\check{p}^\prime_k, \mathbf{p}_{\setminus k}])$, where $\check{p}^\prime_k>\check{p}_k$ is the counterpart power on $u_k([p_k, \mathbf{p}^\prime_{\setminus k}])$ such that $u_k([\check{p}^\prime_k, \mathbf{p}^\prime_{\setminus k}])=u_k([r_k, \mathbf{p}^\prime_{\setminus k}])$, where the inequality $\check{p}^\prime_k>\check{p}_k$ follows from the fact the maximum of $u_k([p_k, \mathbf{p}^\prime_{\setminus k}])$ is placed at $\gamma^\star/\nu^\prime_k>\gamma^\star/\nu_k$. Hence, $u_k([r_k/\Delta, \mathbf{p}_{\setminus k}])<u_k([\check{p}^\prime_k, \mathbf{p}^\prime_{\setminus k}])=u_k([r_k, \mathbf{p}^\prime_{\setminus k}])$, which implies $r_k^\prime=r_k$.
\end{enumerate}

As a conclusion, $r_k(\mathbf{p}_{\setminus k})$ is an ascending best response function. In particular, if \eqref{eq:size} is satisfied, this implies that $r_k(\mathbf{p}_{\setminus k}) \ge r_k(\dot{\mathbf{p}}_{\setminus k})$ for all $k$ and for all $\mathbf{p}$ such that $p_k\ge\dot p_k$, where the vector $\dot{\mathbf{p}}=[\dot p_1, \dots, \dot p_K]^T$ is the minimum component-wise power allocation such that $\gamma_k=\nu_k(\dot{\mathbf{p}}_{\setminus k}) \dot p_k\ge\dot \gamma$ for all $\dot p_k$, with $\dot\gamma<\tilde\gamma$ defined above. In other words, the equilibrium points of $\mathcal{G}$ (if any) are equal to the equilibria of a modified (generalized) game, which differs from $\mathcal{G}$ as now the strategy spaces are subset of $\mathcal{P}_k$ such that any vector allocation $\mathbf{p}$ is such that $p_k\ge\dot p_k$, and hence $\gamma_k\ge\dot\gamma$. To conclude the proof, let us introduce the following definition:
\begin{definition}[\!\!\cite{fudenberg91, topkis79}]\label{def:supermodular}
 A game is \emph{supermodular} if $u_k(\mathbf{p})$ has increasing differences in $\mathbf{p}=[p_k, \mathbf{p}_{\setminus k}]$, i.e., if
 \begin{align}\label{eq:supermodular}
  u_k(\mathbf{p}) - u_k([p^\prime_k, \mathbf{p}_{\setminus k}]) \ge u_k([p_k, \mathbf{p}^\prime_{\setminus k}]) - u_k(\mathbf{p}^\prime)
 \end{align}
 for all $\mathbf{p}$ and $\mathbf{p}^\prime$ such that, for all $k$, $p_k\le p^\prime_k$, and $p_\ell \le p^\prime_\ell$ for all $\ell\neq k$.\hfill$\blacksquare$
\end{definition}

To prove that the utility $u_k(\mathbf{p})$, which is twice differentiable, satisfies \eqref{eq:supermodular}, we can show, using \cite{topkis79}, that possesses the necessary and sufficient condition $\partial^2 u_k(\mathbf{p})/\partial p_\ell \partial p_k \ge0$ for any two components $p_\ell\neq p_k$. Using \eqref{eq:utility}, we can easily derive
\begin{align}\label{eq:utilitySecondDerivative}
 \frac{\partial^2 u_k(\mathbf{p})}{\partial p_\ell \partial p_k}=-\frac{\alpha_\ell}{V\alpha_k}\cdot
 \frac{\Pi^{\prime\prime}_{\textrm{d},k}(\gamma_k) \gamma_k^2-2f(\gamma_k)}{p_k^4}\ge0
\end{align}
where the inequality follows from the fact that $\Pi^{\prime\prime}_{\textrm{d},k}(\gamma_k)=\partial^2 \Pi_{\textrm{d},k}(\gamma_k)/\partial \gamma^2_k\le0$ for $\dot\gamma\le\gamma_k\le\tilde\gamma$, and $f(\gamma_k)=\gamma_k\cdot\partial\Pi_{\textrm{d},k}(\gamma_k)/\partial \gamma_k-\Pi_{\textrm{d},k}(\gamma_k)>0$ for $\gamma_k\le\tilde\gamma$ (see \cite{bacci12a, BSLP12} for further details). Proving the property of supermodularity concludes the proof, as supermodular games admit pure-strategy equilibria \cite{fudenberg91}. As a consequence, the original game $\mathcal{G}$ has pure-strategy GNE, under the sufficient condition \eqref{eq:size}.

\section{Proof of \theoremname~\ref{th:uniqueness}}\label{app:uniqueness}

To show that the GNE of the game $\mathcal{G}$ is not necessarily unique, we use a counterexample. Let us focus on one GNE $\mathbf{p}^\star$, whose existence is ensured by \theoremname~\ref{th:existence}, and let us suppose that $\mathbf{p}^\star$ is such that $p_k^\star\le\gamma^\star/(\nu_k^\prime\sqrt{\Delta})$, where $\nu_k^\prime=\nu_k(\mathbf{p}^\prime_{\setminus k})$ is obtained using the vector $\mathbf{p}^{\prime}_{\setminus k}=\Delta\mathbf{p}^\star$ such that all components are scaled by the quantization step $\Delta$, i.e., $p_k^\prime=\Delta p_k^\star$. Under this hypothesis, $\mathbf{p}^\prime$ is also a GNE of $\mathcal{G}$, i.e., $r_k(p^\prime_{\setminus k})=p_k^\prime$ $\forall k\in\mathcal{K}$.

To show this property, let us note first that $\nu_k^\star<\nu_k^\prime<\nu_k^\star/\Delta$, where $\nu_k^\star=\nu_k(\mathbf{p}_{\setminus k}^\star)$. Hence, if $p_k^\star\le\gamma^\star/(\nu_k^\prime\Delta)$ is the best response $p_k^\star=r_k(\mathbf{p}_{\setminus k}^\star)$, it implies that $p_k^\prime=\Delta p_k^\star\le\gamma^\star/\nu_k^\prime$ is also $p_k^\prime=r_k(\mathbf{p}_{\setminus k}^\prime)$, as $|p_k^\prime-\gamma^\star/\nu_k^\prime|<|p_k^\star-\gamma^\star/\nu_k^\star|$ and $u_k([p_k, \mathbf{p}_{\setminus k}^\prime])$ is an increasing function of $p_k$. If $\gamma^\prime/(\nu_k^\star\Delta)<p_k^\star\le\gamma^\prime/(\nu_k^\star\sqrt{\Delta})$, then $\gamma^\prime/\nu_k^\star<p_k^\prime\le\gamma^\star\sqrt{\Delta}/\nu_k^\star$, i.e., $p_k^\prime=\Delta p_k^\star$ is greater than the point of maximum $\gamma^\prime/\nu_k^\star$. Furthermore, $|p_k^\prime-\gamma^\star/\nu_k^\prime|<|p_k^\star-\gamma^\star/\nu_k^\prime|$. Due to the asymmetry of $u_k([p_k, \mathbf{p}_{\setminus k}^\prime])$, this ensures that $u_k([p^\prime_k, \mathbf{p}_{\setminus k}^\prime])>u_k([p^\star_k, \mathbf{p}_{\setminus k}^\prime])$, and thus $p_k^\prime=r_k(\mathbf{p}_{\setminus}^\prime)$. As a conclusion, $\mathbf{p}^{\prime}=\Delta\mathbf{p}^\star$ is also a GNE of $\mathcal{G}$, and this concludes the proof.

Note that, when $K\gg1$ (e.g., $K=\lfloor1+V/\gamma^\star\rfloor$), it is often the case that $\sum_{\ell\neq k}{\alpha_\ell p_\ell^\star\gg\sigma_n^2}$, $\forall k$. Hence, $\nu_k^\prime\gtrsim\nu_k^\star/\Delta$, and the condition $p_k^\star\le\gamma^\star/(\nu_k^\prime\sqrt{\Delta})$ occurs frequently. This is the reason why the number of GNE increases as $K$ increases. However, such condition is not necessary, and other GNE might exist, e.g., vectors $\mathbf{p}^\prime$ in which some elements are $p^\prime_k=\Delta p_k^\star$ and some others are $p^\prime_k=p_k^\star$.

\section{Proof of \theoremname~\ref{th:PoS}}\label{app:PoS}

By using the relation \eqref{eq:gne}, $u_k(\mathbf{p}^\star_\Delta)=u_k([p_{\Delta,k}^\star, \mathbf{p}^\star_{\Delta,\setminus k}])\ge u_k([p_k, \mathbf{p}^\star_{\Delta,\setminus k}])$ for all $k$ and for all $p_k\in\mathcal{P}_k$. In particular, $u_k([p_{\Delta,k}^\star, \mathbf{p}^\star_{\Delta,\setminus k}])\ge u_k([p_k^\star, \mathbf{p}^\star_{\Delta,\setminus k}])$ for any $p_k^\star$ in any $\mathbf{p}^\star\in{\mbox{$\mathcal{E}$}}^\star$, $\mathbf{p}^\star\neq\mathbf{p}^\star_\Delta$. Note also that $\gamma_k^\prime=\nu_k(\mathbf{p}^\star_{\Delta,\setminus k}) p_k^\star\ge\nu_k(\mathbf{p}^\star_{\setminus k}) p_k^\star=\gamma_k^\star$ for all $k$, and $\gamma_k^\prime>\gamma_k^\star$ for some $k$, under the hypothesis $p_{\Delta,k}^\star\le p_{k}^\star$ for all $k$, $\mathbf{p}_\Delta^\star\neq\mathbf{p}^\star$. As a consequence, $u_k([p^\star_k, \mathbf{p}^\star_{\Delta,\setminus k}])=\Pi_{\mathrm{d},k}(\gamma_k^\prime)/p_k^\star\ge\Pi_{\mathrm{d},k}(\gamma_k^\star)/p_k^\star=u_k(\mathbf{p}^\star)$. Since $u_k(\mathbf{p}^\star_\Delta)\ge u_k(\mathbf{p}^\star)$ for all $k\in\mathcal{K}$, with strict inequality for some $k$, \eqref{eq:so} follows.

\section*{Acknowledgment}
The authors would like to thank Dr. S. M. Perlaza for many helpful discussions and useful comments on draft versions of the manuscript, and L. Marchetti for helpful suggestions on the simulation platform.

\bstctlcite{mybibfile:BSTcontrol}
\bibliography{IEEEabrv,mybibfile}

\newpage

\begin{figure}
 \begin{center}
  \psfrag{xaxis}[c][b]{{$K$}}
  \psfrag{yaxis}[c][t]{{$\textsf{NMSE}(\mathbf{p}_\mathrm{c}^\star)$}}
  \psfrag{data1}[Bl][m]{\!\!\!\footnotesize{$\db[\Delta]=0.5$}}
  \psfrag{data2}[Bl][m]{\!\!\!\footnotesize{$\db[\Delta]=1.0$}}
  \psfrag{data3}[Bl][m]{\!\!\!\footnotesize{$\db[\Delta]=2.0$}}
  \includegraphics[width=0.55\columnwidth]{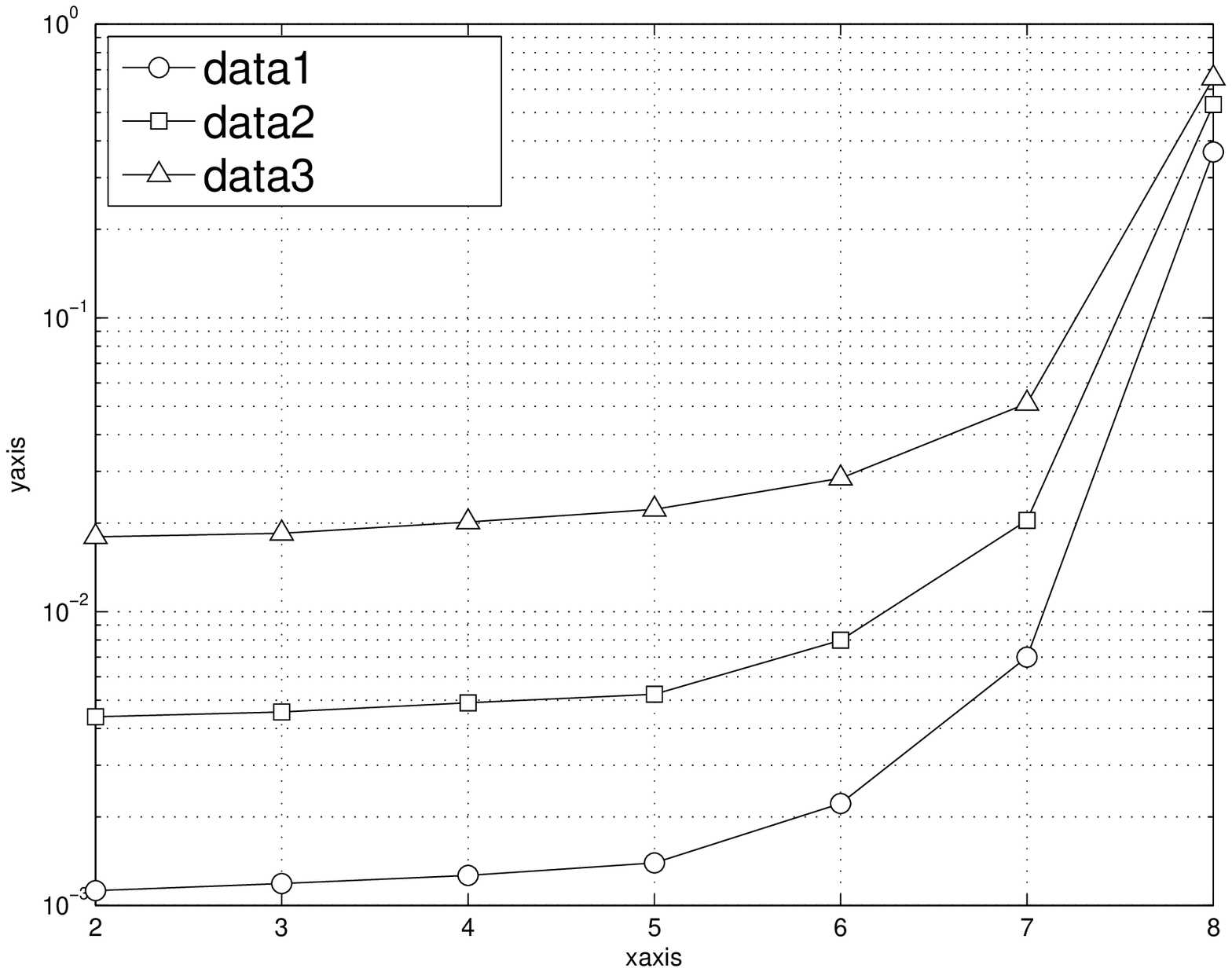}
 \caption{Normalized MSE between power allocations at the GNE for $\mathcal{G}_\mathrm{c}$ and $\mathcal{G}$ as a function of $K$.}
 \label{fig:nmse}
 \end{center}
\end{figure}

\begin{figure}
 \begin{center}
  \psfrag{xaxis}[c][b]{{$K$}}
  \psfrag{yaxis}[c][t]{{$\sum_{k}{u_k(\mathbf{p}_\Delta^\star)}/\sum_{k}{u_k(\mathbf{p}_\mathrm{c}^\star)}$}}
  \psfrag{data1}[Bl][m]{\!\!\!\footnotesize{$\Delta=0.5$ dB}}
  \psfrag{data2}[Bl][m]{\!\!\!\footnotesize{$\Delta=1.0$ dB}}
  \psfrag{data3}[Bl][m]{\!\!\!\footnotesize{$\Delta=2.0$ dB}}
  \includegraphics[width=0.55\columnwidth]{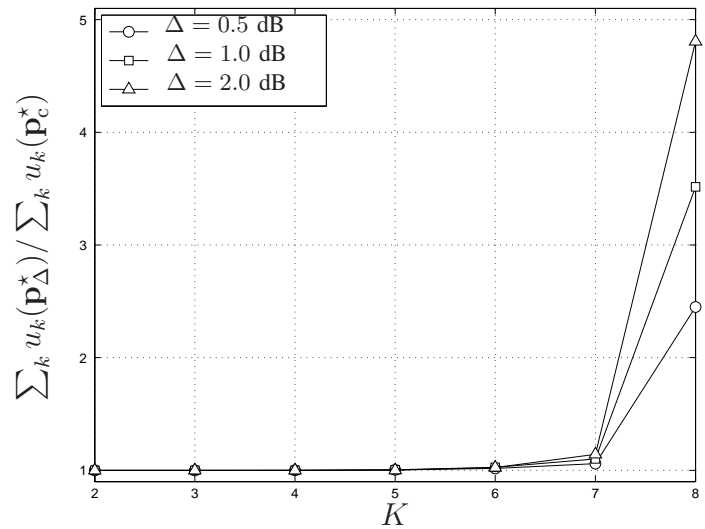}
 \caption{Normalized social welfare increase of $\mathbf{p}_\Delta^\star$ with respect to $\mathbf{p}_\mathrm{c}^\star$ as a function of $K$.}
 \label{fig:utility}
 \end{center}
\end{figure}

\begin{figure}
 \begin{center}
  \psfrag{xaxis}[c][b]{{$K$}}
  \psfrag{yaxis}[c][t]{{$\db[{p_{\textrm{avg}}/\sigma_n^2}]$}}
  \psfrag{data1}[Bl][m]{\!\!\!\!\!\footnotesize{BRSA}}
  \psfrag{data2}[Bl][m]{\!\!\!\!\!\footnotesize{DLF-BRSA ($B\!\rightarrow\!\infty$)}}
  \psfrag{data3}[Bl][m]{\!\!\!\!\!\footnotesize{DLF-BRSA ($B\!=\!8$)}}
  \psfrag{data4}[Bl][m]{\!\!\!\!\!\footnotesize{DLF-BRSA ($B\!=\!3$)}}
  \psfrag{data5}[Bl][m]{\!\!\!\!\!\footnotesize{DLF-BRSA ($B\!=\!2$)}}
  \psfrag{data6}[Bl][m]{\!\!\!\!\!\footnotesize{DLF-BRSA ($B\!=\!1$)}}
  \includegraphics[width=0.55\columnwidth]{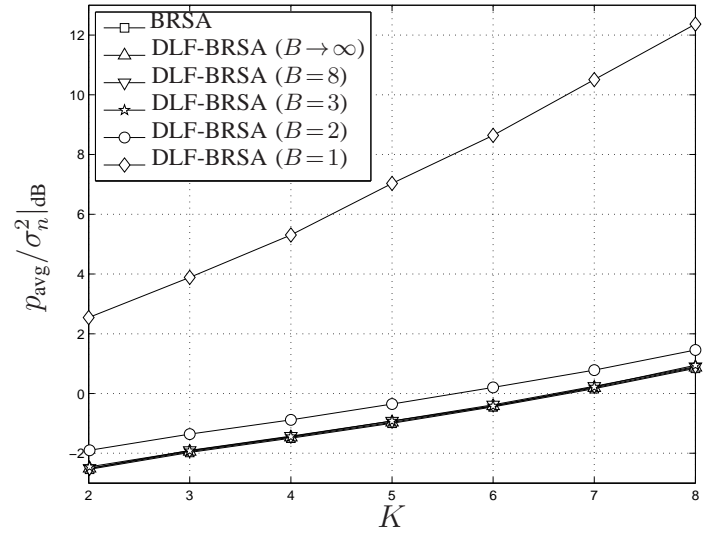}
 \caption{Average power consumption as a function of the number of STs.}
 \label{fig:powerVsUser}
 \end{center}
\end{figure}

\begin{figure}
 \begin{center}
  \psfrag{xaxis}[c][b]{{$K$}}
  \psfrag{yaxis}[c][t]{{$n^{\textrm{exit}}_{\textrm{avg}}$}}
  \psfrag{data1}[Bl][m]{\!\!\!\!\!\footnotesize{BRSA}}
  \psfrag{data2}[Bl][m]{\!\!\!\!\!\footnotesize{DLF-BRSA ($B\!\rightarrow\!\infty$)}}
  \psfrag{data3}[Bl][m]{\!\!\!\!\!\footnotesize{DLF-BRSA ($B\!=\!8$)}}
  \psfrag{data4}[Bl][m]{\!\!\!\!\!\footnotesize{DLF-BRSA ($B\!=\!3$)}}
  \psfrag{data5}[Bl][m]{\!\!\!\!\!\footnotesize{DLF-BRSA ($B\!=\!2$)}}
  \psfrag{data6}[Bl][m]{\!\!\!\!\!\footnotesize{DLF-BRSA ($B\!=\!1$)}}
  \includegraphics[width=0.55\columnwidth]{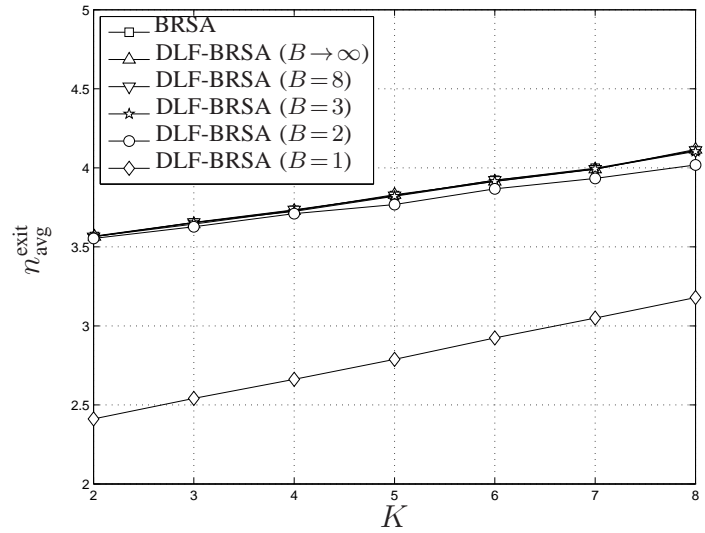}
 \caption{Average number of iterations as a function of the number of STs.}
 \label{fig:timeVsUser}
 \end{center}
\end{figure}

\begin{figure}
 \begin{center}
  \psfrag{xaxis}[c][b]{{$d_1/R$}}
  \psfrag{yaxis}[c][t]{{$\db[{p_{\textrm{avg}}/\sigma_n^2}]$}}
  \psfrag{data1}[Bl][m]{\!\!\!\!\!\footnotesize{DLF-BRSA}}
  \psfrag{data2}[Bl][m]{\!\!\!\!\!\footnotesize{DSA}}
  \psfrag{data3}[Bl][m]{\!\!\!\!\!\footnotesize{BEB-DSA}}
  \includegraphics[width=0.55\columnwidth]{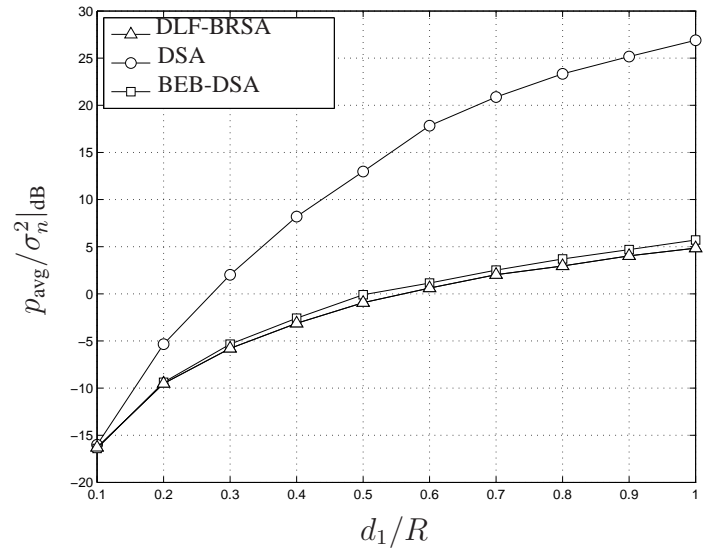}
 \caption{Average power consumption as a function of the normalized distance.}
 \label{fig:powerVsDistance}
 \end{center}
\end{figure}

\begin{figure}
 \begin{center}
  \psfrag{xaxis}[c][b]{{$d_1/R$}}
  \psfrag{yaxis}[c][t]{{$T_{\textrm{avg}}$ [ms]}}
  \psfrag{data1}[Bl][m]{\!\!\!\!\!\footnotesize{DLF-BRSA}}
  \psfrag{data2}[Bl][m]{\!\!\!\!\!\footnotesize{DSA}}
  \psfrag{data3}[Bl][m]{\!\!\!\!\!\footnotesize{BEB-DSA}}
  \includegraphics[width=0.55\columnwidth]{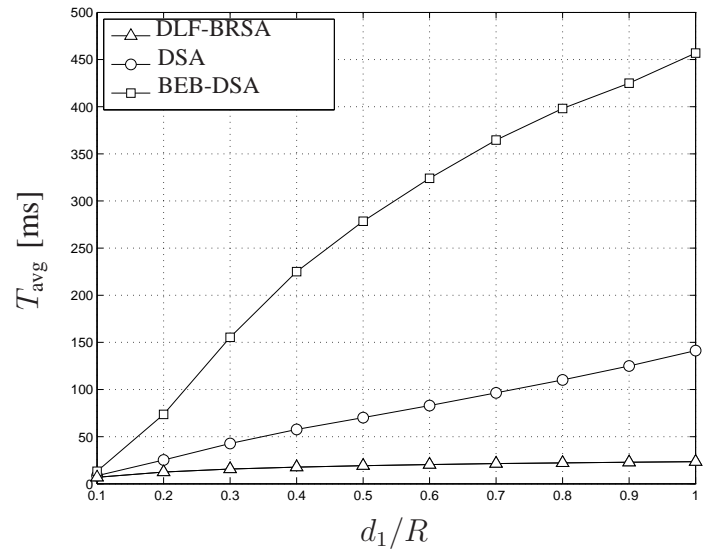}
 \caption{Average synchronization time as a function of the normalized distance.}
 \label{fig:timeVsDistance}
 \end{center}
\end{figure}

\begin{figure}
 \begin{center}
  \psfrag{xaxis}[c][b]{{$d_1/R$}}
  \psfrag{yaxis}[c][t]{{$\mathsf{MSE}(\hat\theta_1)$}}
  \psfrag{data1}[Bl][m]{\!\!\!\!\!\footnotesize{DLF-BRSA}}
  \psfrag{data2}[Bl][m]{\!\!\!\!\!\footnotesize{DSA}}
  \psfrag{data3}[Bl][m]{\!\!\!\!\!\footnotesize{BEB-DSA}}
  \psfrag{data4}[Bl][m]{\!\!\!\!\!\footnotesize{constraint $\overline{\mathsf{MSE}}_\theta$}}
  \includegraphics[width=0.55\columnwidth]{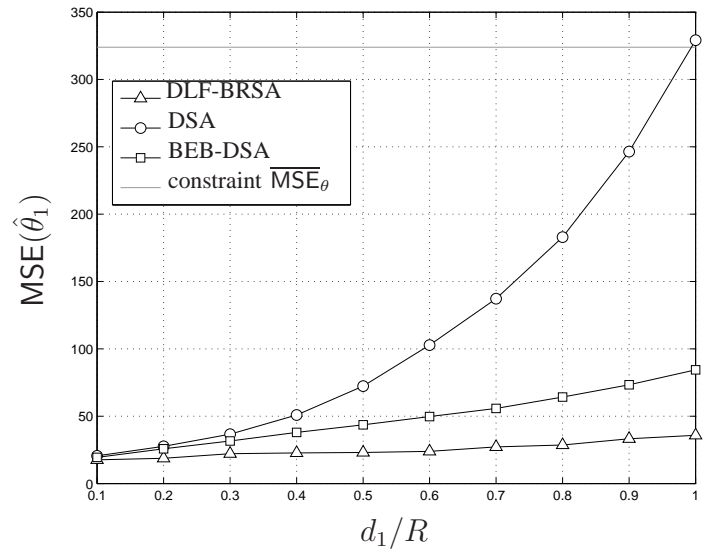}
 \caption{Average timing MSE as a function of the normalized distance.}
 \label{fig:mseVsDistance}
 \end{center}
\end{figure}

\begin{figure}
 \begin{center}
  \psfrag{xaxis}[c][b]{{$p_k$}}
  \psfrag{yaxis}[bc][b]{{$u_k(\mathbf{p})$}}
  \psfrag{a}[c][b]{\scriptsize{$\underline p_k$}}
  \psfrag{b}[c][b][1][180]{\scriptsize{$\gamma_{\mathrm{req}}/\nu_k$}}
  \psfrag{c}[c][b][1][180]{\scriptsize{$\dot\gamma/\nu_k$}}
  \psfrag{d}[c][b]{\scriptsize{$r_k$}}
  \psfrag{e}[c][b][1][180]{\scriptsize{$\tilde\gamma/\nu_k$}}
  \psfrag{f}[c][b]{\scriptsize{$\overline p_k$}}
  \psfrag{g}[l][b]{\scriptsize{$\,\mathcal{P}_k$}}
  \psfrag{h}[l][b]{\scriptsize{$\,{\mathcal{A}}_k(\mathbf{p}_{\setminus k})$}}
  \psfrag{i}[c][b]{\scriptsize{$\Delta$}}
  \includegraphics[width=0.55\columnwidth]{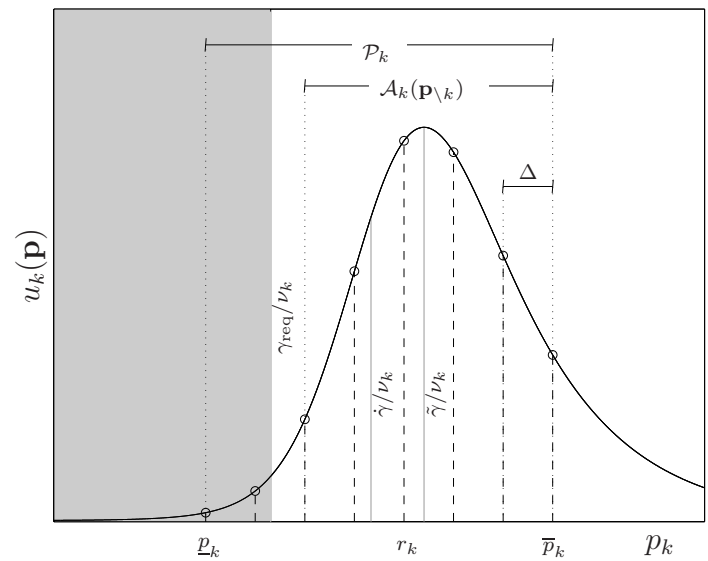}
 \caption{Utility as a function of the transmit power for a fixed interference $\mathbf{p}_{\setminus k}$.}
 \label{fig:existence}
 \end{center}
\end{figure}

\end{document}